\documentclass[aps,twocolumn,prl,showpacs]{revtex4-1}
\usepackage{graphicx}
\usepackage{amsmath}
\usepackage[bf,Large]{}

\begin{document}

\title{Origin of electron-hole asymmetry in the scanning tunneling 
spectrum of $Bi_2Sr_2CaCu_2O_{8+\delta}$} 

\author{Jouko Nieminen$^{1,2}$}\email{jouko.nieminen@tut.fi}

\author{Hsin Lin$^{2}$}\author{R.S. Markiewicz$^{2}$}
\author{A. Bansil$^{2}$}
\affiliation{$^{1}$Institute of Physics,
Tampere University of Technology, P.O. Box 692, 33101 Tampere,
Finland}
\affiliation{$^{2}$Physics Department, Northeastern University, Boston,
Massachusetts 02115}

\begin{abstract}

We have developed a material specific theoretical framework for modelling 
scanning tunneling spectroscopy (STS) of high temperature superconducting 
materials in the normal as well as the superconducting state. Results for 
$Bi_2Sr_2CaCu_2O_{8+\delta}$ (Bi2212) show clearly that the tunneling 
process strongly modifies the STS spectrum from the local density of 
states (LDOS) of the $d_{x^2-y^2}$ orbital of Cu. The dominant tunneling 
channel to the surface Bi involves the $d_{x^2-y^2}$ orbitals of the four 
neighbouring Cu atoms. In accord with experimental observations, the 
computed spectrum displays a remarkable asymmetry between the processes of 
electron injection and extraction, which arises from contributions of Cu 
$d_{z^2}$ and other orbitals to the tunneling current.

\end{abstract}

\date{Version of \today}

\maketitle

Scanning tunneling spectroscopy (STS) has entered the realm of 
high-temperature superconductors powerfully by offering atomic scale 
spatial resolution in combination with high energy resolution. The physics 
of these materials is dominated by the cuprate layers, which are usually 
not exposed to the tip of the apparatus. Much of the existing 
interpretation of the spectra is based however on the assumption that the 
STS spectrum is directly proportional to the
LDOS of the CuO$_ 2$ layer, neglecting the effects of the tunneling 
process in modifying the spectrum in the presence of the insulating 
overlayers. Here, we focus on the Bi2212 system, which has been the 
subject of an overwhelming amount of experimental 
work\cite{McElroy,Hudson,Yazdani,Review,expgap}, although our results bear 
more generally on the STS spectra of the cuprates. 

Our analysis takes into account the fact that the current originating
in the CuO$_{2}$ layers reaches the tip after being `filtered' through
the overlayers of SrO and BiO, and shows that instead of being a
simple reflection of LDOS of the CuO$_{2}$ layers, the STS signal
represents a complex mapping of the electronic structure of the
system. In particular, we find that the spectrum develops a striking
asymmetry between positive and negative biases because $d_{z^2}$ and
other orbitals begin to contribute to the tunneling current with
increasing bias voltage. Although this asymmetry has often been taken
to be the hallmark of strong correlation effects\cite{Hanaguri2004},
our results indicate that the nature of the tunneling process itself
induces significant electron-hole asymmetry even within the
conventional picture, so that strong correlation effects on the STS
spectrum will be more subtle than has been realized so far.

\begin{figure}[th]
\centering \includegraphics[angle=270,width=0.5\textwidth]{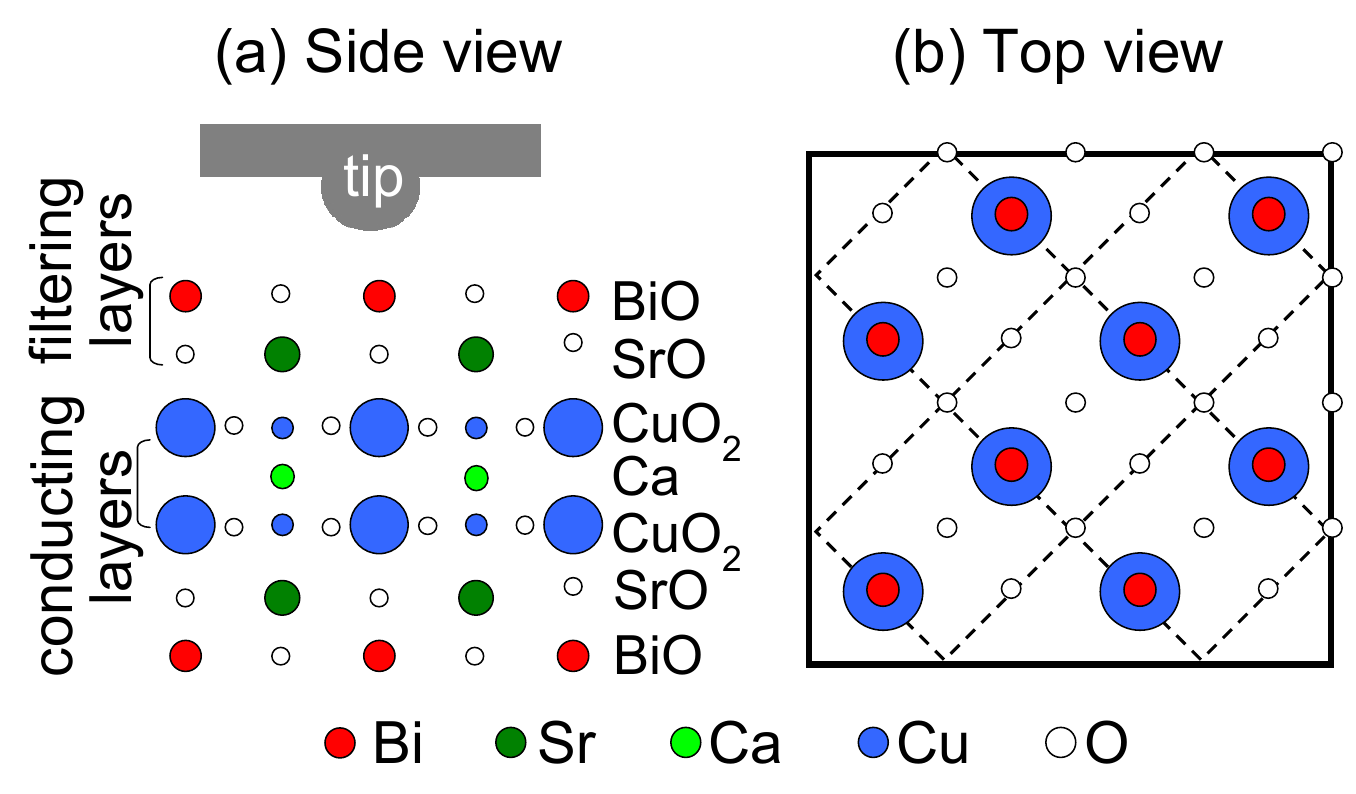}
\caption{(a) Side view of tip placed schematically on top of seven
  layers used to compute the tunneling spectrum of Bi2212, where the
  surface terminates in the BiO layer. Tunneling signal from the
  conducting CuO$_2$ layers reaches the tip after passing through SrO
  and BiO layers. (b) Top view of the surface showing arrangement of
  various atoms. Eight two-dimensional real space primitive unit cells
  used in the computations are marked by dashed lines.}
\label{geometric}
\end{figure}

\begin{figure*}[th]
\centering
  \includegraphics[angle=270,width=1.0\textwidth]{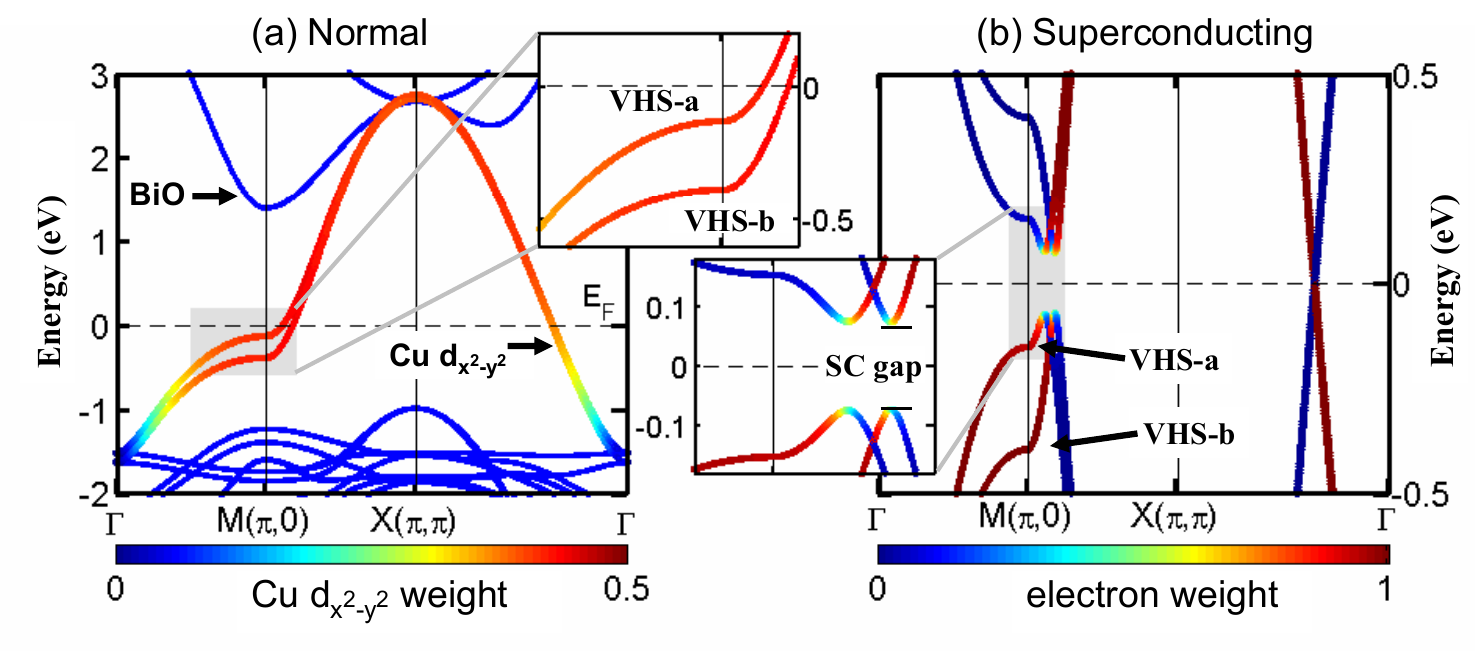}
\caption{
(a) Normal state electronic spectrum of Bi2212. Cu~$d_{x^2-y^2}$
weight of states is given by the color of dots. Inset shows
the energy region of the VHS's on an expanded scale. (b)
Electronic spectrum of the superconducting state. Electron weight
of states is given by the color of dots. Inset highlights the
region of the superconducting gap near the antinodal point.  }
\label{bandsupers}
\end{figure*}

In order to construct a realistic framework capable of describing the
STS spectrum of the normal as well as the superconducting state, we
start with the normal state Hamiltonian
\begin{equation}
\hat{H}_1 = \sum_{\alpha\beta\sigma}
\left[\varepsilon_{\alpha}c^{\dagger}_{\alpha \sigma} c_{\alpha \sigma}+
V_{\alpha \beta}
c^{\dagger}_{\alpha \sigma} c_{\beta\sigma}\right],\label{H1}
\end{equation}
which describes a system of tight-binding orbitals created (or
annihilated) via the real-space operators $c^{\dagger}_{\alpha
  \sigma}$ (or $c_{\alpha \sigma}$). Here $\alpha$ is a composite
index denoting both the type of orbital (e.g. Cu $d_{x^2-y^2}$) and
the site on which this orbital is placed, and $\sigma$ is the spin
index.  $\epsilon_\alpha$ is the on-site energy of the $\alpha^{th}$
orbital.  $\alpha$ and $\beta$ orbitals interact with each other
through the potential $V_{\alpha\beta}$ to create the energy
eigenstates of the system.

Superconductivity is included by adding a pairing
interaction term $\Delta$ in the Hamiltonian:
\begin{equation}
\hat{H} = \hat{H}_1 + \sum_{\alpha \beta
\sigma} \left[\Delta_{\alpha \beta} c^{\dagger}_{\alpha \sigma}
c^{\dagger}_{\beta -\sigma} + \Delta_{\beta \alpha}^{\dagger}
c_{\beta -\sigma} c_{\alpha \sigma} \right]
\label{hamiltonian}
\end{equation}
We take $\Delta$ to be non-zero only between $d_{x^2 - y^2}$ orbitals
of the nearest neighbor Cu atoms, and to possess a d-wave form, i.e.
$\Delta$ is given in momentum space by $ \Delta_k = \frac{\Delta}{2}
\left[\cos{k_x a} - \cos{k_y a} \right],$ where $a$ is the in-plane
lattice constant. This interaction allows electrons of opposite spin
to combine into superconducting pairs such that the resulting
superconducting gap is zero along the nodal directions $k_x=\pm k_y$,
and is maximum along the antinodal directions.

The Bi2212 sample is modeled as a slab of seven layers in which the
topmost layer is BiO, followed by layers of SrO, CuO$_2$, Ca, CuO$_2$,
SrO, and BiO, as shown in Fig. 1(a). The tunneling computations are
based on a $2\sqrt{2} \times 2\sqrt{2}$ real space supercell
consisting of 8 primitive surface cells with a total of 120 atoms (see
Fig.~1(b)). The coordinates are taken from the tetragonal crystal
structure of Ref.~\onlinecite{Bellini}.  
The tip-surface interaction is modeled by Slater-Koster overlap
integrals between an s-orbital with a constant density of states 
and the surface orbitals in the
vicinity of the tip. The tip is allowed to scan across the
substrate to generate the topographic STM map, or held fixed on top of
a surface Bi atom for the STS spectra.

\begin{figure*}[th]
\centering
  \includegraphics[angle=270,width=1.00\textwidth]{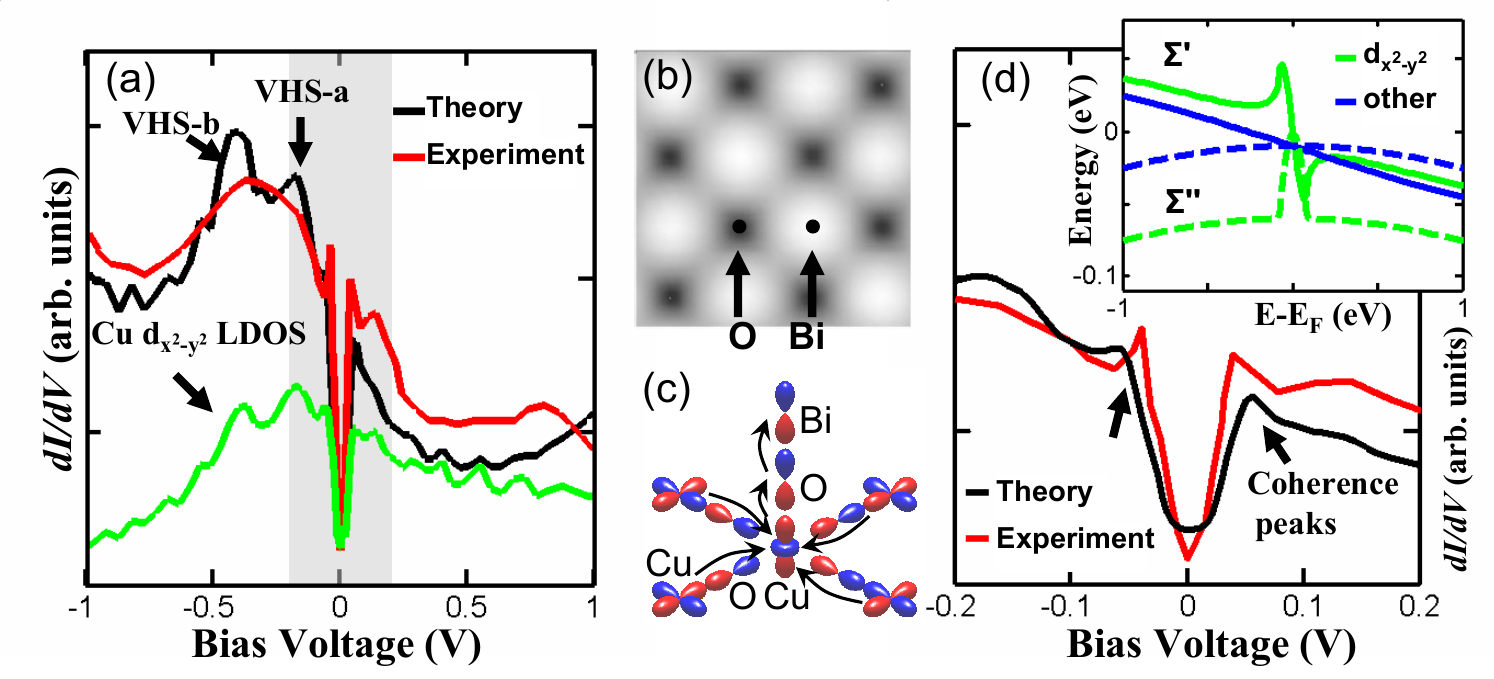}
\caption{ (a) Computed (black line) and experimental (red line)
  differential tunneling spectra, $dI/dV$, as a function of bias
  voltage (in volts). LDOS of Cu~$d_{x^2-y^2}$ electrons is shown
  (green line) as a function of energy, where the horizontal scale is
  in eV.  (b) Computed topographic image of the BiO surface in which
  Bi atoms are bright and O atoms are dark. (c) The main tunneling
  channel from Cu~$d_{x^2-y^2}$ electrons on nearest neighbor Cu atoms
  to the tip through overlap with the central Cu atom, followed by
  hopping to the $p_z$ orbitals of the apical O-atom and the Bi atoms.
  (d) Computed and measured spectra in the low energy region marked by
  gray shading in (a). Inset gives the real (solid lines) and
  imaginary (dashed lines) parts of the self-energy applied to the
  Cu~$d_{x^2-y^2}$ states and to all other states \cite{footnote1}.}
\label{pristine}
\end{figure*}

The tight-binding parameters are fitted to the LDA band structure of
Bi2212 that underlies for example the extensive angle-resolved
photointensity computations of Ref.~\onlinecite{HL}. The Slater-Koster
results\cite{Slater, Harrison} are used to fix the angular dependence
of the tight binding overlap integrals. The specific orbital sets used
for various atoms are: ($s,p_x,p_y,p_z$) for Bi and O; $s$ for Sr; and
($4s,d_{3z^2-r^2},d_{xy},d_{xz},d_{yz}, d_{x^2-y^2}$) for Cu atoms.
This yields 58 orbitals in a primitive cell, used in band
calculations, and a total of $464$ orbitals for Green function
supercell calculations in 256 evenly distributed k-points.  Finally a
gap parameter value of $\vert\Delta\vert = 0.045eV$ is chosen to model
a typical experimental spectrum\cite{McElroy} for the generic purposes
of this study.

The LDOS and tunneling computations are based on Green function
formalism.  At first, the normal-state Green function is constructed
via Dyson's equation using methodology described in
Ref.~\onlinecite{Nieminen}.  At this stage a self-energy for orbital
$\alpha$,
$\Sigma^{\pm}_{\alpha} = \Sigma{'}_{\alpha} \pm i \Sigma{''}_{\alpha}$
is embedded in Dyson's equation for possible effects of various
bosonic couplings and correlation effects \cite{hogenboom,VHS}.  For
simplicity, we have assumed the self-energy to be diagonal in the
chosen basis \cite{footnote1}.
In building up the Green
function in the superconducting state, we utilize the conventional
BCS-type self-energy $\Sigma^{BCS} = \Delta G^{h} \Delta^{\dagger}$
(see, e.g., Ref.~\onlinecite{Fetter}), where $G^{h}$ is the hole part
of normal state Green function.

Fig. 2 shows the calculated band structure of Bi2212 in the normal and
the superconducting state from Hamiltonians of Eqs.~\ref{H1} and
\ref{hamiltonian}. The normal state is seen to properly display the
major features such as: The pair of CuO$_{2}$ bands crossing the Fermi
energy ($E_F$) with the associated van Hove singularities (VHS's)
marked VHS-a (antibonding) and VHS-b (bonding), split by 250~meV at
the $(\pi,0)$ point; BiO bands lying about 1 eV above $E_F$; and the
`spaghetti' of bands involving various Cu and O orbitals starting at a
binding energy of around 1 eV below $E_F$. Although states near $E_F$
are mainly of Cu~$d_{x^2-y^2}$ and O~$p_{x,y}$ character, they also
contain some Bi and Cu~$d_{z^2}$ admixture.  In the superconducting
state in Fig. 2(b), a quasiparticle spectrum mirrored through $E_F$ is
obtained with a doubled number of bands due to the pairing
interaction. A d-wave superconducting gap opens up in both CuO$_2$
bands near $E_F$. The quasiparticles have a mixed electron/hole
character only near the edges of the gap.

To compute the tunneling spectra we apply the Todorov-Pendry
expression \cite{Todorov,Pendry} for the differential conductance
$\sigma$ between orbitals of the tip ($t,t'$) and the sample($s,s'$),
which in our case yields
\begin{equation}
\sigma = \frac{dI}{dV} = \frac{2 \pi e }{ \hbar} \sum_{t t' s s'}
\rho_{tt'}(E_F)V_{t's} \rho_{ss'}^{}(E_F+eV)V_{s't}^{\dagger},
\label{conductance}
\end{equation}
where the density matrix
$\rho_{s s'} = -\frac{1}{\pi}\sum_{\alpha} G_{s \alpha}^{+} \Sigma{''}_{\alpha}
G_{\alpha s'}^{-}$
is, in fact, the spectral function written in terms of retarded/advanced Green
function and the self-energy. Eq.~\ref{conductance} differs from the
more commonly used Tersoff-Hamann approach\cite{Tersoff} in that it
takes into account the details of the symmetry of the tip orbitals and
their overlap with the surface orbitals.

The use of the spectral function recasts Eq.~\ref{conductance} into the form
\begin{equation}
\sigma = \sum_{t \alpha} T_{t \alpha},
\label{transition}
\end{equation}
where
\begin{align}
T_{t \alpha} =& -\frac{2 e }{ \hbar}\sum_{t' s s'}
\rho_{tt'}(E_F)V_{t's}G^{+}_{s\alpha}\Sigma{''}_{\alpha}G^{-}_{\alpha
  s'}V_{s't}^{\dagger},
\label{partial}
\end{align}
and the Green functions and the self-energy are evaluated at energy $E =
E_F + e V_b.$ Eq.~\ref{partial} is similar to the Landauer-B\"uttiker formula
for tunneling across nanostructures (see, e.g., Ref.~\onlinecite{Meir}),
and represents a slight reformulation of Refs.~ \onlinecite{Fisher} and
\onlinecite{Frederiksen}.

The nature of Eq.~\ref{partial} can be understood straightforwardly:
$G_{s\alpha}$ gives the amplitude with which electrons residing on the
$\alpha^{th}$ orbital in the solid propagate to the surface at energy
$E$ broadened by $\Sigma{''}_{\alpha}$. The term $V_{s t}$ is
the overlap between the surface orbital and the tip, while $\rho_{t
t'}$ gives the available states at the tip. Hence, $T_{t \alpha}$
gives the contribution of the $\alpha^{th}$ orbital to the current,
and the summation in Eq.~\ref{transition} collects these individual
contributions to yield the total tunneling current which reaches the
tip. Thus, selecting individual terms in Eq.~\ref{partial} provides a
transparent scheme to define tunneling paths between the sample and
the microscope tip.


Fig.~\ref{pristine}(a) shows the tunneling spectra over the broad
energy range of $\pm$ 1 eV.  At high positive voltages, the computed
spectrum (black line) is fairly structureless. At low energies, a gap
accompanied by the characteristic peak-dip-hump features is observed.
The calculations show the anti-bonding (VHA-a) and bonding (VHS-b)
VHS's\cite{VHS} as distinct structures extending from -0.1 to -0.5 eV, 
followed by a broad dip around
-0.7 eV and subsequent rise near -1 eV \cite{footnote2,Markiewicz}. 
In all these respects,
the present computations substantially follow the experimental results
(red line).\cite{McElroy} In particular, theory reproduces the
observed asymmetry of the tunneling spectrum with excess intensity at
negative biases.  The rapid increase in current at high binding
energies results from increasing spectral weight of Cu~$d_{z^2}$ and
other orbitals contributing to the `spaghetti' of bands starting
around 1 eV binding energy (see Fig.  2(a)). We emphasize that the
LDOS of the Cu~$d_{x^2-y^2}$ (green line in Fig. 3(a)) does not
provide a good description of the spectrum. In particular, the
Cu~$d_{x^2-y^2}$ LDOS possesses an asymmetry which is opposite to that
of the tunneling spectrum.

Fig. 3(d) gives a blow up of the low energy region of $\pm 0.2 eV$,
shown by gray shading in Fig. 3(a). The computed spectrum is seen to
reproduce the coherence peaks and the characteristic peak-dip-hump
feature. The generic form of the real and imaginary parts of the
self-energies applied to the Cu $d_{x^2-y^2}$ orbitals (solid and
dashed blue lines, respectively) and the rest of the orbitals are
given in the inset \cite{footnote1}. Fig.~3(b) shows the computed
`topographic map' of the BiO surface in constant current mode.  Bi
atoms appear as bright spots in accord with experimental observations,
while O~atoms sit at the centers of dark regions.

\begin{figure}[th]
\includegraphics[angle=0,width=0.5\textwidth]{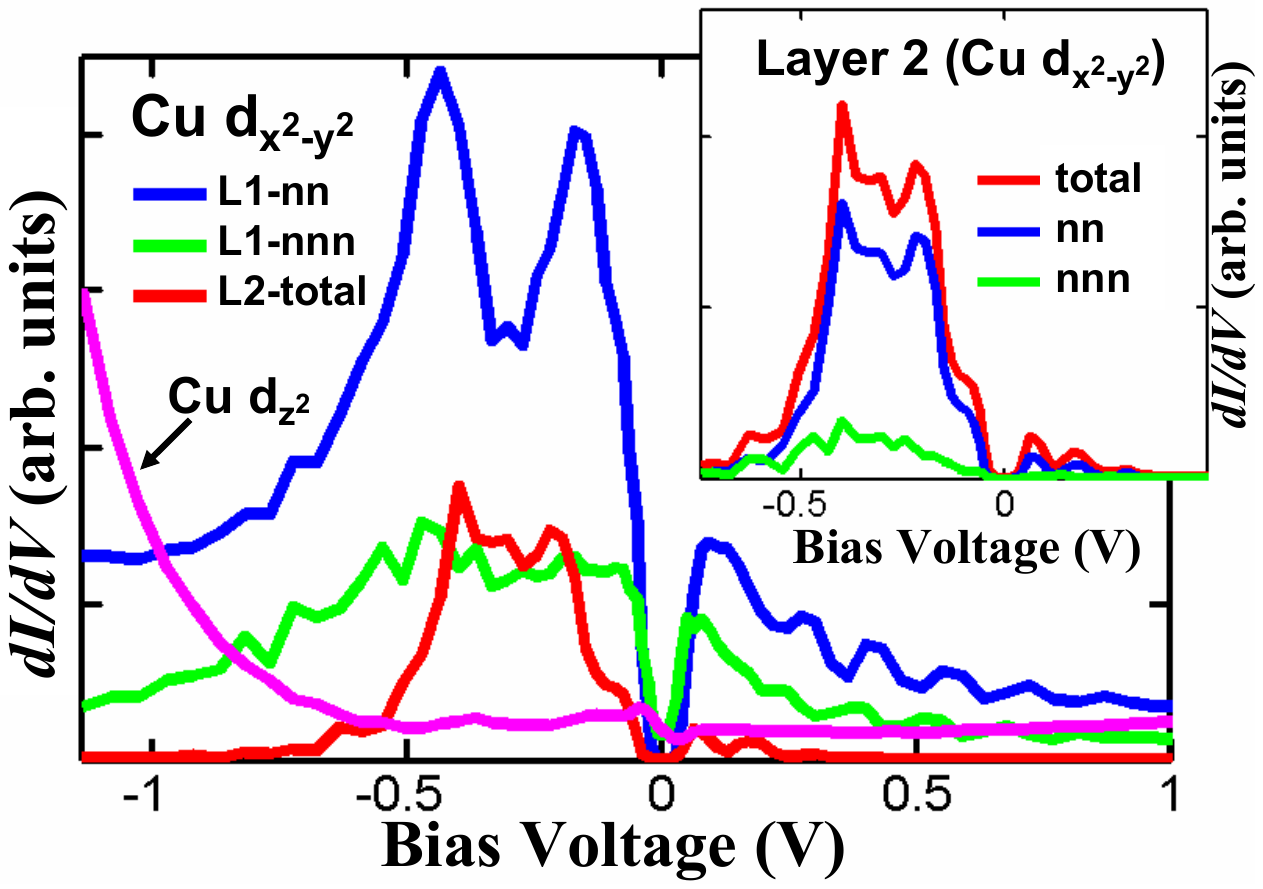}
\caption
{{\it Main frame}: Partial contributions to the tunneling current from
  various orbitals in the two cuprate layers. The CuO$_2$ layer
  closest to the tip is identified as layer 1 or L1, while the second
  layer is denoted by L2. Specific contributions are: $d_{x^2-y^2}$
  orbitals of the four nearest neighbor Cu atoms (L1-nn, blue line);
  $d_{x^2-y^2}$ orbitals of the four next nearest neighbor Cu atoms of
  the first layer (L1-nnn, green line) ; $d_{x^2-y^2}$ orbitals of the
  Cu atoms of the second layer (L2, red line); $d_{z^2}$ of the
  central Cu atom of L1 (magenta line). {\it Inset}: Decomposition of
  the current from the second cuprate layer: Total contribution (red
  line); contribution of the four nearest neighbours (blue line); and
  the next nearest neighbours (green line).}
\label{channels}
\end{figure}

An analysis of the partial terms of Eq.~\ref{partial} reveals that the
$d_{x^2-y^2}$ orbital of the Cu atom lying right under the Bi atom
gives zero contribution to the current \cite{Balatsky}. The dominant
contribution to the spectrum comes from the four nearest neighbor (NN)
Cu atoms as indicated schematically in Fig.~3(c).  A detailed
decomposition of Eq.~\ref{channels} is shown in Fig.~\ref{channels},
where paths starting from the CuO$_2$ layer closest to the tip (L1),
as well as from the second cuprate layer (L2) are considered.  The
signal from cuprate layers is dominated by the $d_{x^2-y^2}$ orbitals
on the four nearest neighbour (nn) Cu atoms in L1 up to about -0.7 eV
(blue line).  At higher binding energies, the contribution from the
$d_{z^2}$ electrons from the Cu atom right below the Bi atom or the
tip grows rapidly (magenta line).

A smaller but still significant contribution comes from the four next
nearest neighbour (nnn) $d_{x^2-y^2}$ orbitals in L1 spread over a
wide energy range (green line, main figure), while the total current
originating from the $d_{x^2-y^2}$ orbitals of L2 is quite localized
over zero to -0.6 eV bias (red line, main figure).
Fig.~\ref{channels} emphasizes the nature of the current associated
with the cuprate layers and points out an intrinsic electron-hole
asymmetry originating from the $d_{z^2}$ orbitals. We note however
that the Bi and O orbitals in the surface Bi-O layer can also play a
role in producing an asymmetric background current.

In conclusion, we find that STS spectrum for Bi2212 is strongly
modified from the LDOS of $d_{x^2-y^2}$ by the effect of the tunneling
process or what we may call the tunneling matrix element. Much of the
observed asymmetry of the spectrum can be explained within the
conventional picture due to the turning on of Cu~$d_{z^2}$ and other
channels with increasing (negative) bias voltage. This indicates that
the effects of strong electronic correlations on the tunneling
spectrum are more subtle than has been thought previously.  However,
we should note that we have not analyzed spectra associated with the
deeply underdoped regime where charge order has been reported
\cite{Kohsaka}. The present method naturally allows an analysis of the
tunneling signal in terms of the possible tunneling channels and the
related selection rules. Our scheme can be extended to incorporate
effects of impurities and various nanoscale inhomogeneities by using
appropriately larger basis sets in the computations.

{\bf Acknowledgments}

This work is supported by the US Department of Energy, Office of
Science, Basic Energy Sciences contract DE-FG02-07ER46352, and
benefited from the allocation of supercomputer time at NERSC,
Northeastern University's Advanced Scientific Computation Center
(ASCC), and the Institute of Advanced Computing, Tampere.


\begin{thebibliography}{99}
\bibitem{footnote1}
The specific self-energy used is shown in the inset to Fig. 3(d). This 
includes a Fermi liquid broadening ($E^2$) of all the levels. The 
peak-dip-hump feature in the $d_{x^2-y^2}$ orbitals is modeled via 
coupling to a generic bosonic mode.
\bibitem{footnote2}
The distinct VHS peaks in the computed spectrum are expected to be
broadened to yield a broad hump much like the experimental spectrum
due to self-energy corrections resulting from magnetic response of the
electron gas in the -400 meV range (see, e.g., Ref.~\onlinecite{Markiewicz}). 
These self-energy corrections are not included in
the present calculations.
\end{thebibliography}
\end{document}